\documentclass[aps,pra,floatfix,showpacs,noshowkeys,epsfig,graphics]{revtex4}%
\usepackage{graphicx}
\usepackage{amsmath}
\usepackage{amsfonts}
\usepackage{amssymb}

\newcommand{\newc}{\newcommand}
\newc{\beq}    {\begin{equation}}
\newc{\eeq}    {\end{equation}}
\newc{\beqa}    {\begin{eqnarray}}
\newc{\eeqa}    {\end{eqnarray}}
\newc{\bs}    {\section}
\newc{\no}    {\\ \nonumber}

\begin{document}
%\title{ Absence of Entanglement in Thermal Spin One Boson Systems.}
\title{ Quantum Separability of Thermal Spin One Boson Systems.}
\author{Jae-Weon Lee}\email{scikid@kias.re.kr}
\author{Sangchul Oh}\email{scoh@kias.re.kr}
\author{Jaewan Kim}\email{jaewan@kias.re.kr}
\affiliation{School of Computational Sciences,
Korea Institute for Advanced Study,
             207-43 Cheongnyangni 2-dong, Dongdaemun-gu,
              Seoul 130-012, Korea}
\date{\today}
\begin{abstract}
Using the temperature Green's function approach we investigate entanglement
between  two non-interacting spin 1 bosons in  thermal
equilibrium. We show that, contrary to the fermion case, the entanglement is absent
in the spin density matrix.
Separability is demonstrated using the Peres-Horodecki criterion
for massless particles such as photons in black body radiation.
For massive particles, we show that the density matrix can be decomposed with
separable states.

\end{abstract}
%\keywords{entanglement; solid state qubits; quantum information}
\maketitle
Recent progress in  quantum information theories and experiments \cite{Nielsen01,Vedral02,Galindo02}
has led to interests in studying the non-locality
and entanglement in many particle systems\cite{
O'Connor01,Arnesen01,Wang01,Osborne02,Osterloh02,Vidal03,Glaser03,Schliemann01,Eckert02,Paskauskas01,Wiseman03,%
Omar02,Gittings02} such as Bose Einstein condensations\cite{bec}, Heisenberg Model\cite{spin},
Fermion systems\cite{Vedral03,oh}, and superconductors\cite{ohsc} and
even in the vacuum\cite{reznik}.
Entanglement is now treated as a physical quantity
 like energy and entropy,
as well as a resource for quantum information processing.
 Spin 1 particles such as photons and $W^{\pm}$ and $Z^0$ gauge vector bosons
are essential ingredients in the standard model. Furthermore, black body radiation(BBR),
 the historical birth place of the
quantum physics,  still plays an important role in many
fields such as quantum optics\cite{qoptics}, the black hole radiation\cite{hawking} and the
cosmic microwave background radiation\cite{cbr}. Thus, studying
entanglement of thermal spin 1 bosons is important.

In this letter, we use the  temperature Green's function approach to
investigate the  quantum entanglement of two non-interacting
(massless and massive ) spin 1 boson particles in  thermal
equilibrium.
 Vedral\cite{Vedral03}
studied the entanglement in many body systems at zero temperature
using the second quantization formalism. Following his works, Oh
and Kim\cite{oh} studied the entanglement of  two electron spins
in a free electron gas, superconductivity\cite{ohsc} and the Kondo model\cite{kondo} at finite
temperature using thermal Green's function methods.
 In his work, Vedral
showed that there is no reason why the polarizations of a pair of separated
photons should be correlated at zero
temperature. At finite temperature, however, the situation becomes more
complicated. In this case
entanglement may occur because, contrary to the intuition that thermal
noise destroys entanglement, it has been  shown  that even if two particles
do not interact directly, they can become entangled by interacting
with a common heat bath\cite{braun,rie}.
  To maintain thermal equilibrium, the thermal bosons (like  photons in the BBR)
should interact with a common thermal bath, even when they
are not directly interacting  with each other.
 Therefore we need to calculate the entanglement of thermal spin 1 bosons
explicitly to check for the absence of entanglement in the systems.
Recently, it was also shown that two qubits interacting with the
BBR can be entangled\cite{brauncbr}.
In earlier works entanglement in many body systems was usually tested indirectly by
investigating the entanglement of two`probe qubits' interacting with the system.
In this letter, however, we are
interested in the entanglement of the particles themselves without
any probe qubit. This approach
could reveal the physical nature of the system more clearly.

We begin by briefly reviewing the Green's function approach\cite{Mahan}.
 To calculate the entanglement we need to know the
density matrix of the system with Hamiltonian $H$ and temperature $1/\beta$.
 The finite temperature two-particle density matrix is defined with the field operator $\hat{\psi}(x_i)$
 for the $i$-th particle;
\begin{eqnarray}
\rho^{(2)}(x_1,x_2;x_1'x_2') &\equiv& \frac{1}{2} \langle
\hat{\psi}^{\dag}(x_2') \hat{\psi}^{\dag}(x_1')
                             \hat{\psi}(x_1)\hat{\psi}(x_2) \rangle \no
                              &=& -\frac{1}{2}
   {\cal G}(x_1\tau_1, x_2\tau_2 ; x_1'\tau_1'^{+}, x_2'\tau_2'^{+})\,
\end{eqnarray}
where  $\langle {\cal O}\rangle = {\rm Tr}\{\rho_G {\cal O}\}$
with $Z ={\rm Tr}\{ e^{-\beta H}\}$ and $\rho_G= e^{-\beta H }/Z$,
 and $\tau^{+}_i~(i=1,2)$ denotes a time infinitesimally later than $\tau_i$.
Using the Wick's theorem,
the two-particle temperature Green's function  for bosons
 can be reduced to the product of  one-particle Green's
functions;
\begin{eqnarray}
 \label{Hatree-Fock}
{\cal G}(1,2;1',2') \equiv
&\text{Tr}&\{ \hat{\rho}_{G}\, T_{\tau}[
      \hat{\psi}_{K}(1)
      \hat{\psi}_{K}(2)
      \hat{\psi}_{K}^{\dag}(2')
      \hat{\psi}_{K}^{\dag}(1')]\}\no
      &\approx& {\cal G}(1,1'){\cal G}(2,2') + {\cal
G}(1,2'){\cal G}(2,1')\,
\end{eqnarray}
where the number $i~(i=1,2)$ denotes the space-time coordinates $(x_i,\tau_i)$ of particle
$i$, and ${\cal G}(1;1') \equiv \text{Tr}\{ \hat{\rho}_{G}\, T_{\tau}[
\hat{\psi}_{K}(1) \hat{\psi}_{K}^{\dag}(1')]\}$ is the
one-particle temperature Green's function.
 The field operator is redefined as $\hat{\psi}_{K}(x,\tau) =
e^{\hat{K}\tau/\hbar}\,\hat{\psi}(x)\,e^{-\hat{K}\tau/\hbar}$ with
$\hat{K} = \hat{H} -\mu\hat{N}$, where
 $\mu$ is the chemical potential and $\hat{N}$ is the number operator.
The second equality of Eq.~(\ref{Hatree-Fock}) denotes the Hartree-Fock approximation
which is exact
for non-interacting systems such as the one considered in this paper.
Then,
the non-interacting one particle Green's function ${\cal G}^0(1;1')$ is
\begin{eqnarray}
\rho^{(1)}(x;x') = -{\cal G}^{0}(x\tau;x'\tau^{+}) =
\delta_{\sigma\sigma'}\, g({\bf r} -{\bf r}') ,
\label{Eq:one-density}
\end{eqnarray}
where $\sigma$ denotes the spin index and  $g({\bf r}-{\bf r}')$
is  the one-particle space density matrix in a volume $V$;
\begin{eqnarray}
g({\bf r}) = \frac{1}{V}\sum_{{\bf k}} e^{i{\bf k}\cdot{\bf r}}
n_{\bf k}.
\end{eqnarray}
Here $n_{\bf k}= \{\exp[\beta(\epsilon_{\bf k} - \mu)] - 1\}^{-1}$
is the mean occupation number in the state with momentum $\bf k$
and energy $\epsilon_{\bf k} = \hbar^2k^2/2m$  for massive
non-relativistic bosons or $\epsilon_{\bf k} = \hbar k/c$ for
photons. With Eqs. (\ref{Hatree-Fock}) and (\ref{Eq:one-density}),
one has the explicit form for the two-particle space-spin density
matrix~\cite{Yang62,Loewdin55}
\begin{eqnarray}
\rho^{(2)}(x_1,x_2;x_1',x_2')
= \frac{1}{2}[
   g({\bf r}_1- {\bf r}_1')g({\bf r}_2- {\bf r}_2')
  \delta_{\sigma_1\sigma_1'}\delta_{\sigma_2\sigma_2'}
 + g({\bf r}_1- {\bf r}_2')g({\bf r}_2- {\bf r}_1')
   \delta_{\sigma_1\sigma_2'}\delta_{\sigma_1'\sigma_2}]\,.\qquad
\end{eqnarray}
where $\sigma_i$ denotes the spin index for the $i$-th particle.
To the best of our knowledge, there is still no consensus on how
to deal with entanglement between  continuous
variables such as coordinates and discrete variables such as spin. Hence we set
${\bf r}_1 =
{\bf r}_1'$ and ${\bf r}_2 = {\bf r}_2'$
 to consider only discrete (spin) degrees of freedom,
 which leads to a simpler form for the density matrix.
For isotropic cases, the two-spin
density matrix, depending on the relative distance between two
particles $r = |{\bf r}_1 - {\bf r}_2|$, is
\begin{eqnarray}
\rho^{(2)}_{\sigma_1,\sigma_2;\sigma_1'\sigma_2'}(r) =
\frac{n^2}{2 \alpha^2}\left[
    \delta_{\sigma_1\sigma_1'}\delta_{\sigma_2\sigma_2'}
 + f(r)^2 \delta_{\sigma_1\sigma_2'}\delta_{\sigma_1'\sigma_2}\right]\,,
\label{Eq:two_spin_matrix}
\end{eqnarray}
where $\alpha$ is the number of spin degrees of the freedom
($\alpha=2$ for
massless spin 1 bosons and 3 for massive ones). $n \equiv  N/V$ is the particle
 density
for particle number $N$ and
$f(r)$ is an exchange term representing the indistinguishability of bosons:
\begin{eqnarray}
f(r) \equiv \frac{\alpha}{n}g({\bf r})
      = \frac{\alpha}{N}\sum_{{\bf k}} e^{i{\bf k}\cdot{\bf r}} n_{\bf k} ,
\label{f}
\end{eqnarray}
A bipartite state $\rho$ is called separable  if it can be written in the form
\beq
\rho=\sum_{i=1}^n p_i \rho^A_{i}\otimes \rho^B_{i},
\label{separable}
\eeq
where $\rho^A_{i}$ and $\rho^B_{i}$ are states of subsystem $A$ and $B$, respectively.
We use the Peres-Horodecki separability criterion\cite{peres,Horodecki96},
which is the positive partial transpose(PPT) criterion.
A state is PPT if $\rho^{T_B}>0$, where the partial transposition of $\rho$ is
\beq
\rho^{T_B}_{im,jn}\equiv \langle i,m|\rho^{T_B}|j,n\rangle= \rho_{in,jm}
\eeq
in some basis.
 Let us first
consider massless particles such as photons from BBR with
  two spin degrees of freedom
 denoted by two
level states$(|0\rangle,|1\rangle)$. The two-spin density matrix corresponds to
the two qubits density matrix
in this case.
 By dividing the bracket part
of Eq.~(\ref{Eq:two_spin_matrix}) by $4+2f^2$, we obtain the
normalized two-spin density matrix $\rho_{12}$ \cite{Yang62} for a
given relative distance $r$ between two photons in $\{ |00\rangle,
|01\rangle,|10\rangle,|11\rangle\}$ polarization basis
\begin{eqnarray}
\rho_{12} = \frac{1}{4+2f^2}
            \begin{bmatrix}
            1 + f^2 & 0 & 0  &0 \\
            0       & 1 & f^2 &0 \\
            0       &f^2 & 1  &0 \\
            0       & 0 & 0 &1+f^2
           \end{bmatrix}\,,
\end{eqnarray}
where $\text{Tr}_{\sigma_1\sigma_2}\{\rho_{12}\} = 1$ and we have dropped $r$ in $f(r)$ for simplicity.
 This matrix has the same form as for the Fermion case
except that the off-diagonal terms have plus signs\cite{oh}. One can
easily show that $\rho_{12}$ is PPT and hence separable\cite{ifonlyif}
(The lowest eigenvalue of $\rho_{12}^{T_B}$is $1/(4 + 2f^2)>0$).
Hence we can conclude that there is no entanglement in the two-spin
density matrix of the non-interacting massless thermal spin 1 boson system.

What can the absence of entanglement in the BBR
be used for
from a practical viewpoint? The
absence of quantum correlation can
help us  to understand the
nature of certain light sources, for example, astronomical objects.
By performing
an Aspect-type Bell test experiment\cite{aspect} on polarization states of two photons from a light source
and checking for violation of the Bell inequality\cite{chsh},
 one could determine whether the source emits entangled
photons.
If this test reveals
 pairs of entangled photons from the source,
 one can say that the source is, at least, not a black-body radiator.
 Given that information from distant astronomical objects is mainly obtained by observing electromagnetic waves,
this quantum test would provide  us additional useful information about the objects.

We now move on to the case of  massive spin 1 particles,
which have 3 spin states ($\alpha=3$).
In this case Eq.(\ref{Eq:two_spin_matrix}) reads, in $\{|00\rangle,|01\rangle,|02\rangle,|10\rangle,\cdots,
|22\rangle\}$ basis,
\begin{eqnarray}
\rho_{12} = \frac{1}{9+3f^2}
  \begin{bmatrix}
 1 + f^2 & 0 & 0 & 0 & 0 & 0 & 0 & 0 & 0 \ \cr
 0 & 1 & 0 & f^2 &0 & 0 & 0 & 0 & 0 \cr
 0 & 0 & 1 & 0 & 0 & 0 & f^2 & 0 & \ 0\cr
  0 & f^2 & 0 & 1 & 0 & 0 & 0 & 0 & 0 \cr
  0 & 0 & 0 & 0 & 1 +f^2 & 0 & 0 & 0 & 0 \cr
   0 & 0 & 0 & 0 \ & 0 & 1 & 0 & f^2 & 0 \cr
    0 & 0 & f^2 & 0 & 0 & 0 & 1 & 0 & 0 \cr
0 & 0 & 0 & \ 0 & 0 & f^2 & 0 & 1 & 0 \cr
0 & 0 & 0 & 0 & 0 & 0 & 0 & 0 & 1 + f^2 \cr
 \end{bmatrix}
\end{eqnarray}
,which describes a two-qutrit mixed state. Now
$\rho_{12}^{T_B}$ has the smallest eigenvalue $1/(9 + 3f^2)>0$,and hence $\rho_{12}$ is PPT.
Since PPT is a necessary, but not a sufficient, condtion\cite{bound} for separability in a 3$\times$3 system,
PPT alone does not automatically guarantee separability.
We therefore need to show  explicitly  that there is no bound entanglement in $\rho_{12}$.
Indeed, it can be shown that $\rho_{12}$ can be
  decomposed  with separable states (i. e., in the form of Eq.(\ref{separable}))
  as follows;
\beq
\rho_{12}= \frac{1}{9 + 3 f^2}(9  f^2 \rho_0 + 3\sigma_0  +
        6(1 - f^2)\sigma_1 ),
\eeq where
\beqa \sigma_0&\equiv&\frac{1}{3}(|00\rangle \langle 00|
+|11\rangle \langle 11|+|22\rangle \langle 22|) ~~\text{and} \no
\sigma_1&\equiv&\frac{1}{6}(|01\rangle \langle 01| +|02\rangle \langle
02|+|10\rangle \langle 10| +|12\rangle \langle 12| +|20\rangle
\langle 20|+|21\rangle \langle 21|) \eeqa are trivially separable.
Exploiting roots of unity, $\rho_0$ can be written as an integral over  product states; \beq
\rho_0\equiv\int^{2\pi}_{0} \frac{d\theta}{2\pi} |\psi(\theta)\rangle
\langle \psi(\theta)| \otimes |\psi(\theta)\rangle \langle
\psi(\theta)|, \eeq where
$|\psi(\theta)\rangle=\frac{1}{\sqrt{3}}(|0\rangle +
e^{-i\theta}|1\rangle +e^{2i\theta}|2\rangle$).\\
The massive spin 1
boson class contains vector mesons and massive gauge bosons such
as $W^{\pm}$and $Z^0$\cite{SM}, which intermediate forces between subatomic
particles.  They are believed to have been in thermal states in the early universe.
 Another interesting example is spinor Bose Einstein condensation with
 alkalis with hyperfine
spin 1 such as $^{23}Na$ and $^{39}K$ \cite{spinorBEC}. Since
these particles usually have self-interactions, it might be
interesting to study how self-interactions change  our results
(Our results correspond to tree-level approximations in these
cases).

It is well known that, in algebraic quantum field theory, due to the Reeh-Schlieder theorem\cite{reeh}
 the vacuum
for  quantum fields could have quantum
nonlocality\cite{werner,narnhofer}.
Furthermore, in Ref. \cite{narnhofer} entanglements of  Bose (and Fermi) quasi-free (gaussian) states
are extensively studied.
Our results do not contradict
these results, because in our model we consider only the discrete (spin) degree of freedom of two
separated particles in a thermal system.
 In our approach separability of the  two-spin density matrix
is shown explicitly using the familiar Green's function method,
especially for the $3\times 3$ thermal vector boson system.

Several points can be made in
relation to the material presented here.
First, using the Green's function
method we showed that there is no entanglement in the two-spin
density matrix of a non-interacting thermal spin 1 bosons, regardless of their masses.
However our results do not  rule
out the possibility  of
more sophisticated entanglements in these systems,
 such as those between continuous variables and discrete variables\cite{relativity}.
How do two probe qubits interacting with BBR become entangled\cite{brauncbr}, even though the
spin states of two photons in BBR are separable?
One possible explanation would be that BBR has a more
 sophisticated entanglement, as described above.
Another explanation would be that the coupling between two probe
qubits and the common environment (like BBR) induces an  indirect
interaction between the pair of qubits\cite{ohint}, which produces
the entanglement \cite{brauncbr}. Second, the absence of
entanglement of two  photon spins can be useful for
characterization of light sources by checking  the validity of the
Bell inequality experimentally. Finally, it will be interesting to
investigate how self-interactions or higher spins of bosons
changes the results.

P.S. After submission of our paper we found that there appears a
paper about spatial entanglement of free thermal bosonic fields
(quant-ph/0607069).

\paragraph*{Acknowledgments.--}
\indent J. Lee was supported by part by the Korea Ministry of Science and Technology.
S.Oh was partially supported by R\&D Program
for Fusion Strategy of Advanced Technologies of Ministry of
Science and Technology of Korean Government.
J. Kim
was supported by the Korea Research Foundation (Grant
No. KRF-2002-070-C00029).

\end{document}